\documentclass[english]{article}
\usepackage{times}
\usepackage[T1]{fontenc}
\usepackage[latin1]{inputenc}
\usepackage{geometry}
\usepackage{rotating}
\usepackage{color}
\usepackage{graphicx}
\usepackage{setspace}
\usepackage{amssymb}
\usepackage[authoryear]{natbib}

\usepackage{babel}
\makeatother

\makeatletter

\title{Supplementary Information}
\begin{document}
\date{}
\maketitle

\section{Derivation of Diffusion Equations}
\label{derivation}
In this technical section, we construct the Kolmogorov equations which determine the dynamics of the probability distribution function $P(x,t)$.  In order to do this, we first calculate the transition probabilities between the various states $x\in \{0, \frac{1}{N},\frac{2}{N}, ... ,1\}$

Let $T_\uparrow(x)$ denote the probability that the system makes a transition from the state with a fraction $x$ of mutators to the state with a fraction $x+\frac{1}{N}$ of mutators.  This may occur in one of the following two ways:
\begin{enumerate}
\item A mutator is selected for birth, a wild-type is selected for death, and no mutation occurs.
\item A mutator is selected for birth, a wild-type is selected for death, a beneficial mutation occurs, and this mutation is part of the fraction $1-s$ that is destined for loss by random drift .
\end{enumerate}
\noindent
Computing these probabilities in the order listed, we arrive at the following expression for $T_\uparrow(x)$
\begin{eqnarray}
\nonumber \frac{T_\uparrow(x)}{r} =& &x(1-x)(1-\mu_+) + x(1-x)\mu_+\alpha_e (1-s)\\
= & & x(1-x)\left[1-\mu_+\left(1-\alpha_e (1-s)\right)\right]
\label{hop_plus}
\end{eqnarray}
\noindent
The factor of $r$ on the LHS is just the birth probability per time-step which, according to A1-A3 is common to all members of the population and will soon be scaled out.   
In a similar way we calculate $T_\downarrow(x)$, the probability that the system makes a transition from the state with a fraction $x$ mutators to the state with a fraction $x-\frac{1}{N}$ mutators.  In fact, we may simply interchange $x \leftrightarrow 1-x$ and $\mu_+ \leftrightarrow \mu_-$ in Eq.\ref{hop_plus} which results in
\begin{equation}
\frac{T_\downarrow(x)}{r} =x(1-x)\left[1-\mu_-\left(1-\alpha_e (1-s)\right)\right]\label{hop_minus}
\end{equation}
Within the framework of A1-A3, the population may also make large, non-local  transitions to the ``absorbing'' $x=0$ and $x=1$ states if the mutator or wild-type strains produce an advantageous mutant which is marked for fixation.  This gives rise to 
\begin {eqnarray}
\frac{T_{\textit{fix}}}{r} &=& x\mu_+\alpha_e s \label{trans_fix}\\ 
\frac{T_{\textit{loss}}}{r} &=& (1-x)\mu_-\alpha_e s\label{trans_loss}
\end{eqnarray}
The probability that the population undergoes no change during a timestep is simply what remains
\begin{equation}
\frac{T_o}{r} = 1 - T_\downarrow(x) - T_\uparrow(x) - T_{\textit{fix}} - T_{\textit{loss}}
\end{equation}

These transition probabilities allow us to write down the so called forward and backward Kolmogorov diffusion equations which describe the time dependent probability density $P(x,t)$ that the mutator frequency is $x$ at time $t$.  The forward equation reads:
\begin{eqnarray}
\nonumber \frac{\Delta P(x,t)}{\Delta t} =& -&\left[T_\downarrow(x)+ T_\uparrow(x) \right] P(x,t)\\
\nonumber & + &  T_\downarrow(x+\frac{1}{N})P(x+\frac{1}{N},t) + T_\uparrow(x-\frac{1}{N})P(x-\frac{1}{N},t)\\
& - &\left[T_{fix}(x)+T_{loss}(x)\right]P(x,t)
\label{full_master_eq}
\end{eqnarray}
Taking the continuum limit and plugging in the specific expressions for transition probabilities, we obtain for the forward equation
\begin{eqnarray}
\nonumber \frac{\partial P}{\partial t} &=&\frac{1}{N}\frac{\partial^2}{\partial x^2}\left[x(1-x)P\right]\\
\nonumber &+& \left[1-\alpha_e(1-s)\right](\mu_+-\mu_-)\frac{\partial}{\partial x}\left[x(1-x)P\right] \\
&-& N\alpha_e s\left[x\mu_++(1-x)\mu_-\right]P
\label{full_fpe}
\end{eqnarray}  
where $t$ has been rescaled by $N/r$ so that the units are now ``generations.''  This is Eq(4) in the main text.
 
An approximation to a limited version of Eq.\ref{full_fpe} is solved in section \ref{forward_solution}.  However, we can write an equivalent ``backward Kolmogorov'' equation which is often more mathematically convenient than Eq.  \ref{full_fpe}.   Defining $G(x_o,t)$ as the probability that the mutator has been lost by time $t$, we find 
\begin{equation}
G(x_o,t+\Delta t) = T_\downarrow G(x_o-\frac{1}{N},t) + T_\uparrow G(x_o + \frac{1}{N},t) + T_oG(x_o,t) +T_{loss}(x_o)
\label{discrete_back_eq}
\end{equation}
The backward equation is primarily useful in its steady state form.  Defining $G(x_o,t\rightarrow \infty) \equiv G_\infty(x_o)$ and taking the continuum limit, we obtain the ODE
\begin{eqnarray}
 \nonumber0 &=& \frac{1}{N}\frac{d^2}{dx_o^2}G_\infty \\
 \nonumber  &-&(\mu_+-\mu_-)\left[1-\alpha_e(1-s)\right]\frac{d}{dx_o}G_\infty\\
    &-& N\mu_+\alpha_e s\frac{G_\infty}{1-x_o} +N\mu_-\alpha_e s \frac{1-G_\infty}{x_o}
\label{back_eq}
\end{eqnarray}
This is Eq(6) from the main text.

\section{Limiting Solutions to Eq.\ref{back_eq}, when $\mu_-=0$}
\label{solutions}

As in the main text, we define $B \equiv \mu_+ \left[1-\alpha_e(1-s)\right]$ and $C\equiv \mu_+ \alpha_e s$.  If $N\alpha_e s \gg 1$ but $\mu_+$ is sufficiently small, $NS_\mu$ is no longer much larger than 1, and the approximations in the main text are not valid.  This occurs when $\mu_+ \sim O(1/N^2\alpha_e s)$.  In this case, the $B$ term, and hence deleterious mutations, in Eq.\ref{back_eq} is irrelevant, and $G_\infty(x_o)$ can be expressed in terms of a modified Bessel function:
\begin{equation}
G_\infty(x_o)=\frac{\sqrt{1-x_o}I_1(2N\sqrt{C(1-x_o)})}{I_1(2N\sqrt{C})}
\label{bessel_appx}
\end{equation}
When $N\sqrt{C}$ is not large, this does not have the exponential dependence on $Nx_o$ required to interpret the fixation probability as resulting from a true effective selection coefficient. We can nevertheless
calculate the fixation probability for small $x_o$:
\begin{equation}
P_{fix}(x_o) \approx N\sqrt{C}x_o\frac{I_0(2N\sqrt{C})}{I_1(2N\sqrt{C})}
=N\sqrt{\mu_+\alpha_e s}x_o\frac{I_0(2N\sqrt{\mu_+\alpha_e s})}{I_1(2N\sqrt{\mu_+\alpha_e s})}
\end{equation}
For $\mu_+ \gg 1/(N^2 \alpha_e s)$, the argument of the Bessel function is large, and
we recover our previous result: $P_{fix} \approx Nx_o \sqrt{\mu_+ \alpha_e s}$.  For small
argument, we get $P_{fix}\approx x_o( 1 + N^2C/2) = x_0 ( 1 + N^2\mu_+\alpha_e s/2)$.  Thus
the fixation probability approaches the neutral result $x_o$ as $\mu_+ \to 0$ and
starts out rising linearly in $\mu_+$.  If we wanted to translate this into an effective selection coefficient,
since for small $Ns$, $P_{fix}(x_o)\approx x_o(1+Ns/2)$, the effective selection coefficient would be
$S_\mu = N\mu_+\alpha_e s$, whose explicit $N$ dependence again points to the
inability to define an effective selection coefficient in this regime.

When $N \mu_+  \sim O(1)$ and $N^2\mu_+ \alpha_e s \sim O(1)$, all the terms in the equation are of the same order, and
no approximation can be made.  However, for smaller $\mu_+$, one can use perturbation theory to find an approximate solution by writing $G_\infty = 1-x_o + \eta(x_o)$, where $\eta(x_o) \ll 1-x_o$.  After dropping terms $\sim NB\eta'$ and $\sim N^2C\eta$, we obtain
\begin{equation}
G_\infty(x_o) \approx 1-x_o - \frac{CN - B}{2}Nx_o(1-x_o)
\end{equation}
with a fixation probability $P_{fix}(x_o)\approx x_o(1 + N(CN-B)/2)= x_o[1 + \mu_+N (\alpha_e(Ns + 1) - 1)]$, which linearly approaches the neutral value $x_o$ as $\mu_+ \to 0$.  As above, in this very small $\mu_+$ regime, no mapping to an $N$-independent effective selection coefficient can be made. Note that we again recover our threshold criterion for mutators to be favored (main text Eq (7)).

\section{Approximate Solution to Forward Equation when $\mu_-=0$}
\label{forward_solution}
Eq.\ref{full_fpe} can be approximately solved if we take $\mu_- = 0$.  The equation then reads
\begin{equation}
\frac{\partial P}{\partial t} =\frac{1}{N}\frac{\partial^2}{\partial x^2}\left[x(1-x)P\right]
 + B\frac{\partial}{\partial x}\left[x(1-x)P\right] 
- NCx\mu_+P
\label{appendix_full_fpe}
\end{equation}  
The biological problem we are interested in solving is the fixation probability for a small initial fraction of mutators.  This corresponds to solving for $\int_{1-\epsilon}^{1+\epsilon} P(x,t\rightarrow \infty) dx$ as $\epsilon \rightarrow 0$, subject to the initial condition $P(x,0) = \delta(x-x_o)$, where $x_o \ll 1$ and $\delta(x-x_o)$ is a Dirac delta function.  Furthermore analytic progress can be made if we note that $x$ is in some sense small.  The idea is that the probability cloud $P(x,t)$ is initially localized around $x_o \ll 1$, and that the only process that moves probability solidly into the interior of $x\in (0,1)$ is random genetic drift.  We anticipate this effect to be small when the mutator is significantly favored, i.e. $NS_\mu \gg 1$, and hence $P(x,t) \approx 0$ for $x$ not $\ll 1$.  Thus, we can approximately neglect the $O(x^2)$ terms in Eq.\ref{appendix_full_fpe} and obtain 
\begin{equation}
\frac{\partial P}{\partial t} =\frac{1}{N}\frac{\partial^2}{\partial x^2}\left[xP\right]
 + B\frac{\partial}{\partial x}\left[xP\right] 
- NCx\mu_+P
\label{appendix_appx_fpe}
\end{equation}
This second order PDE in $(x,t)$ can be converted to a first order PDE in $(k,t)$ by taking the spatial Fourier transform, which yields
\begin{eqnarray}
N\frac{\partial \tilde{P}}{\partial t}& = &-i(k^2 -iB k +C)\frac{\partial \tilde{P}}{\partial k} \label{appx_ft}\\ \nonumber
\nonumber  \tilde{P}(k,t=0)& = &\exp{(-ikx_o)} \label {ft_ic}
\end{eqnarray} 

\noindent
This equation can be solved by the ``method of characteristics'', in which we seek curves in the $kt$ plane along which $\tilde{P}(k,t)$ is constant.  We find $\frac{d\tilde{P}}{dt}=\frac{\partial \tilde{P}}{\partial t}+\frac{\partial \tilde{P}}{\partial k}\frac{dk}{dt} = 0$ along the family of curves defined by
\begin{eqnarray}
 \frac{t}{N}&+&\frac{i}{z_+-z_-}\left[\ln \frac{k-z_+}{k-z_-} - \ln \frac{\kappa - z_+}{\kappa - z_-}\right]=0  \label{char_curves} \\ \nonumber \\
 \nonumber z_\pm &\equiv& \frac{iNB}{2}\left[1 \pm \sqrt{1+\frac{4C}{B^2}} \right]
\end{eqnarray}

\noindent
$\kappa$ serves to label different characteristic curves and is chosen to appear in this manner so that $\kappa = k$ when $t=0$.  Then, $\tilde{P}(k,t) = \tilde{P}(k,0) = \tilde{P}(\kappa,0) = \exp{(-i\kappa x_o)}$ along the characteristic curves, and we obtain the formal solution
\begin{equation}
P(x,t) = \frac{1}{2\pi}\int_{-\infty}^{\infty}e^{-i\kappa(k,t)x_o}e^{ikx}dk
\label{formal_soln}
\end{equation}
where $\kappa(k,t)$ is obtained from Eq.\ref{char_curves}.  

This formidable inversion integral gives the full solution for all $x$ and $t$, but fortunately we do not need to evaluate the integral in order to obtain the fixation probability of the mutator.  A moment's reflection convinces us that the $t\rightarrow \infty$ behavior of Eq.\ref{appendix_appx_fpe} is the build-up of a delta function at the absorbing state $x=0$ and a ``decay'' of the remaining probability to the fixation state.  We note that the probability which corresponds to the delta function is the $k\rightarrow \infty$ component of $\tilde{P}(k,t)$.  Taking the $k\rightarrow \infty$ limit of Eq.\ref{char_curves}, we obtain
\begin{eqnarray}
\nonumber P(x=0,t) = e^{-i\kappa_\infty x_o} \\ \nonumber \\
\nonumber \kappa_\infty = z_-\frac{\frac{z_+}{z_-}-e^{-i(z_+-z_-)t/N}}{1 - e^{-i(z_+-z_-)t/N}}
\end{eqnarray}
Finally, taking the $t\rightarrow \infty$ limit and setting $P(1,t\rightarrow \infty) = 1-P(0,t\rightarrow \infty)$, we obtain the familiar expression
\begin{equation}
P(1,t\rightarrow \infty) = 1-e^{x_o |z_- |}\equiv 1-e^{-Nx_oz}
\label{appendix_appx_pfix}
\end{equation}

\begin{equation}
S_\mu  = z =\frac{\sqrt{B^2 + 4C}-B}{2} \approx \frac{\mu_+}{2}\left[\sqrt{(1-\alpha_e)^2 + 4\alpha_e s/\mu_+}-(1-\alpha_e)\right]\quad\quad\quad NS_\mu \gg 1
\label{appendix_appx_S_mu}
\end{equation}
which is the same as Eq(6, main text) obtained from Eq.\ref{back_eq}.

\section{Perturbative Approach to the Effect of $\mu_-$}
The small effect of mutations in wild-type backgrounds observed in simulations motivates a perturbative solution to Eq.\ref{back_eq}.  In terms of the parameters $B_\pm \equiv \mu_\pm [1-\alpha_e(1-s)]$ and $C_\pm \equiv \mu_\pm \alpha_e s$,   
\begin{eqnarray}
 \nonumber   \frac{d^2}{dx_o^2}G_\infty 
 -N(B_+ - B_-)\frac{d}{dx_o}G_\infty
    - N^2C_+\frac{G_\infty}{1-x_o} =-N^2C_- \frac{1-G_\infty}{x_o}
 \end{eqnarray}
In order to make analytic progress, we make the following assumptions.  (i) The mutator is strongly favored, and therefore $\frac{G_\infty}{1-x_o}\rightarrow G_\infty$.  (ii) $G_\infty \approx G_o + G_1$, where $G_o$ is given by the solution to the case $\mu_- = 0$ and $G_o \gg G_1$.  Then we have
\begin{eqnarray}
 G_1''(x_o) 
 -NB_+G_1'(x_o)
  - N^2C_+G_1(x_o) =-N^2C_- \frac{1 - e^{N(B_+-\sqrt{B_+^2+4C_+})x_o/2}}{x_o}
  \label{lo_pert_mum}
  \end{eqnarray}
where we have also dropped the small term $B_- G_1(x_o)$. 
This equation can be solved using the theory of non-homogeneous linear differential equations.  A convenient way to write the two independent solutions to the homogeneous version of Eq.\ref{lo_pert_mum} is 
\begin{eqnarray}
g_<(x_o) &=& e^{B_+Nx_o/2}\sinh\left(\frac{N}{2}\sqrt{B_+^2+4C_+}x_o\right)\nonumber\\
g_>(x_o) &=& e^{B_+Nx_o/2}\sinh\left(\frac{N}{2}\sqrt{B_+^2+4C_+}(1-x_o)\right)\nonumber
\end{eqnarray}
If we denote the inhomogeneity $m(x_o)$, our solution for $G_1(x_o)$ can be written in terms of the integrals
\begin{eqnarray}
G_1(x_o) = \int_0^{x_o}  m(x)\frac{g_<(x)g_>(x_o)}{\textit{Wr}(x)}dx + \int_{x_o}^1 m(x)\frac{g_>(x)g_<(x_o)}{\textit{Wr}(x)}dx \nonumber
\end{eqnarray}
where the Wronskian $\textit{Wr(x)}=g'_>(x)g_<(x)-g_>(x)g'_<(x)$.  The first-order contribution to the fixation probability for small $x_o$ is then
\begin{eqnarray}
 \nonumber F_1(x_o) \approx -x_o \left.\frac{d}{dx_o}G_1(x_o)\right|_{x_o=0} = - x_o\int_0^1 m(x) \frac{g_>(x)g'_<(0)}{\textit{Wr}(x)}dx 
\end{eqnarray}
The Wronskian is evaluated as
\begin{eqnarray}
 \nonumber \textit{Wr}(x)=-\frac{1}{2}e^{B_+Nx}\frac{N}{2}\sqrt{B_+^2+4C_+}\sinh \left(\frac{N}{2}\sqrt{B_+^2+4C_+} \right) 
\end{eqnarray}
Thus, $f_>(x)/\textit{Wr}(x)$ decays rapidly for large $x$ as $e^{-N(B+\sqrt{B_+^2+4C_+})x/2}$.  This allows us to simplify the integral by extending the range of integration to infinity, which yields
\begin{eqnarray}
\nonumber F_1(x_o) \approx-\mu_-\alpha_e sN^2 x_o \int_0^\infty dx \frac{1-e^{N(B_+-\sqrt{B_+^2+4C_+})x/2}}{x} e^{-N(B_++\sqrt{B_+^2+4C_+})x/2}
\end{eqnarray}
Using the identity
\begin{eqnarray}
\nonumber \int_0^\infty dx \frac{e^{-ax}-e^{-bx}}{x}=\ln(b/a)
\end{eqnarray}
we finally arrive at
\begin{eqnarray}
F_1(x_o) \approx- \mu_-\alpha_e s N^2 x_0 \ln\left(\frac{2\sqrt{1+4\frac{\alpha_e s}{\mu_+(1-\alpha_e)} } }{1+\sqrt{1+4 \frac{\alpha_e s}{\mu_+(1+\alpha_e)}}}\right)
\label{pert_correction}
\end{eqnarray}
The logarithmic factor varies between zero in the limit $\mu_+ \gg 4 \alpha_e s$ and $\ln(2)$ in the opposite limit.  This method breaks down when$F_1 \gtrsim F_o$.  Now, $F_o$ is bounded from above by $Nx_o S_\mu^* < Nx_o \alpha_e s$, as given in Eq(11, main text).  Therefore,  Eq.\ref{pert_correction} will typically fail when $\mu_- \alpha_e s N^2 \sim N \alpha_e s$, or, $N\mu_- \sim 1$, which is, unfortunately, usually the case.  

\section{$N_e$ for a Population of Periodically Changing Size}
Whereas our model describes a population of constant size, experiments by \cite{sniegowski1997ehm} were done according to a serial dilution protocol in which a population of size $N_o\approx 5 \times 10^6$ was grown to size $N_f\approx 5 \times 10^8$, diluted 100 fold, then repeated.  Under these dynamics, all lineages grow essentially deterministically from $N_o$ to $N_f$, at which point binomial sampling abruptly reduces the population size back to $N_o$.  In this case, the fixation probability $\pi$ of an advantageous mutant  depends not only on $s$, but also on \textit{when} it is generated during the dilution cycle.  Mutants that are generated during the early part of the cycle are allowed more time to grow exponentially faster than the wild-type and thus have an advantage over late occurring mutants.  It can be shown \citep{wahl55pbm,wahl2002eip} that the stochastic effects of these population bottlenecks are in many ways equivalent to those of a population with constant size $N_e$.  More precisely, if we let $m \equiv$ the number of newly generated mutants that will achieve fixation, then we require that the average value of $\frac{dm}{dt}$ to be the same in the two populations.  In the bottleneck population, the total number of newly generated individuals $\equiv \nu(t) = N_o(e^{t\ln2}-1) $,  and $dm = \mu \pi(s,t) d\nu = N_o\mu  \pi(s,t) \ln(2)e^{t\ln2}dt$.  In the constant size population, $\frac{dm}{dt} = N_e\mu s$.  Equating these two expressions for $\frac{dm}{dt}$ and averaging over one dilution cycle, we obtain 
\begin{equation}
N_e s = \frac{N_o\ln2}{g}\int_0^g e^{t\ln2} \pi(s,t) dt 
\label{define_Ne}
\end{equation}
where $g = \frac{1}{\ln2}\ln(\frac{N_f}{N_o}) \approx 6.6$ is the number of growth generations separating $N_o$ and $N_f$.  For $gs\ln2 \ll 1$ it can be shown \citep{wahl55pbm} that $\pi(s,t) \approx 2s\ln(2)ge^{-t\ln2}$, and therefore Eq.\ref{define_Ne} implies that $N_e = 2N_o g \ln^22\approx 6.3\times 10^7$. 

\section{Detailed Comparison to Experiment}
\begin{table}

\caption{ Values of relevant parameters for non-mutators in \textit{E. coli}, as reported in various references.  We assume that all mutation rates are $100\times$ greater in mutators.  Mutation rates are per genome per replication.  ``Selection coefficient'' refers to that of advantageous mutations only. }

\begin{center}\begin{tabular}{lcccc}
\hline 
Reference & $\mu_{ben}$ & $\mu_{del}$ & U & Selection Coefficient
\tabularnewline
\hline
\cite{hegreness2006epi}&$2.0\times10^{-7}$&                                     &  &.054
\tabularnewline

\cite{lenski1991lte}       &$2.8\times 10^{-10}$&                                  &  &.10
\tabularnewline


\cite{perfeito2007amb} &$2\times 10^{-8}$&                                         & &.023
\tabularnewline

\cite{imhof2001fea}        &$4\times 10^{-9}$&                                          & &.02
\tabularnewline

\cite{rozen2002fef}         &$5.9\times 10^{-8}$ &                                      & &.0235
\tabularnewline

\cite{kibota1996egm}&                                        &$1.9 \times 10^{-4}$& &
\tabularnewline

\cite{keightley1999tma}&                                   &$1.6 \times 10^{-3}$& &
\tabularnewline

\cite{taddei1997mop}&                                    &                                   &$5\times 10^{-7}$&
\tabularnewline

\cite{bo_mutator}&                                             &                                   &$5\times 10^{-6}$&
\tabularnewline
\hline
\end{tabular}\end{center}
\label{param_table}
\end{table}

In biological populations, mutants with a spectrum of beneficial effects are generated at specific rates $\mu_{bp}\rho(s)ds$, where $\rho(s)$ is likely a decreasing function of $s$ \citep{orr2003dfe,eyrewalker2007dfe}.  The weakest mutants are generated frequently, but are unlikely to achieve fixation because (i) their intrinsic fixation probability $\pi \sim s$ is small, and, (ii) in reasonably large populations, several of these mutations exist simultaneously and thus compete with one another.  Conversely, stronger mutants are seldom generated, but likely achieve fixation.  These conflicting influences result in beneficial mutations of some intermediate size $\tilde{s}[\rho(s),N,\mu_{bp}]$ typically achieving fixation \citep{gerrish1998fcb,d_fisher_short,hegreness2006epi}.  These mutants are generated at a per capita rate $\mu_{ben}\approx \mu_{bp}\int_{\tilde{s}}^\infty \rho(s)ds$.  Thus, whenever the population size is large enough for the aforementioned effects to play a strong role, the microscopic parameters $\mu_{bp}$ and $\rho(s)$ result in the \textit{macroscopic parameters} $\tilde{s}$ and $\mu_{ben}$.  These are the parameters that we list in table 2 and plug into our model.  This macroscopic viewpoint tightens the connection between our simple model and experimental reality.  

Plugging in in various parameters from table 2 in to ISLA (see main text), we obtain values of $P_{fix}$ in the range
\begin{equation}
3.5\times 10^{-9} \leqslant P_{fix,isla} \leqslant 1.0 \times 10^{-4} \nonumber
\end{equation}
This range for $P_{fix,isla}$ is strikingly broad, and results from a correspondingly broad range in the beneficial mutation rate.  This rate depends on the particular strain of \textit{E. coli} used, the environmental conditions, the population size \citep{gerrish1998fcb,perfeito2007amb}, and exactly which mutations are counted in calculating the beneficial mutation rate.

\section{Numerical Integration}
In order to produce the solid curves in Figs.(4, 5, 7, 8) from the main text, we first had to numerically integrate Eq.\ref{back_eq}, subject to the boundary conditions $G_\infty(0)=1$ and $G_\infty(1)=0$.  The procedure for the case $\mu_-=0$ is relatively simple.  We initiate integration near the singular point at $x_o=1$, taking $G'_\infty(1-\epsilon) = -1$ and $G_\infty(1-\epsilon) = \epsilon$.  Here, $\epsilon$ is a very small positive number and the initial slope $-1$ is arbitrary.  The integration is then performed from $x_o=1-\epsilon$ to $x_o=0$ using a fourth order Runge-Kutta algorithm.  The resulting trial solution to Eq.\ref{back_eq} does not obey the boundary condition at $x_o=0$.  However, because the equation is linear, the correct solution is obtained simply by re-scaling the trial solution so that the boundary condition is satisfied.  We then evaluate $G_\infty(.001)$ using a cubic spline and obtain $S_\mu$ by inverting Eq(2, main text) using a root solver.  

For $\mu_- > 0$, the procedure is slightly more involved.  Eq.\ref{back_eq} now has singular points at both $x_o=0$ and $x_o = 1$.  Therefore, we must integrate from both the right and the left, then match these two solutions and their derivatives in the middle.  Specifically, we first integrate Eq.\ref{back_eq} from the right, as before, but now stopping at $x_o=.5$.  Call this un-scaled solution solution $G_r(x_o)$.  We then generate a trial solution $G_l(x_o)$ initialized near $x_o = 0$, taking $G_l'(\epsilon) = -NS_o$ and $G_l(x_o) = 1-NS_o \epsilon$.  Here, $S_o$ is given by Eq(10, main text) and merely serves as an initial guess as to the behavior of the solution near $x_o = 0$.  We can ensure that $G_r(.5)=G_l(.5)$ simply by re-scaling $G_r(x_o)$.  However, the slopes will, in general, not match at $x_o=.5$.  In order to accomplish this matching, we link the above procedure to a root solver which repeatedly adjusts $G_l'(\epsilon)$ and generates trial solutions until one is found for which $G'_l(.5)=G'_r(.5)$.  We then proceed to calculate $S_\mu$ as before, using the correct solution $G_l(x_o)$.

\section{Ensemble Averaging}
The point-like symbols in the figures in the main text result from values of $P_{fix}(N,x_o,s,\alpha,\mu_\pm)$ obtained by simulating numerous competition experiments.  The averaging procedure varied somewhat, depending on parameters used, though this had no effect on our results.  Here, we explicitly report the averaging details for each case.

\begin{itemize}
\item All data from populations of size $N=5000$ result from $10,000$ trials run for each $x_o \in \{.003, .009, .015, .021\}$.  The $P_{fix}$ obtained from each value of $x_o$ was then translated into a value for $S_\mu$ via Eq.(2, main text).  These four values were averaged to obtain the values presented in the figures.  
\item For data from populations of size $N=1000$, the procedure was identical to the case where $N=5000$, but with $100,000$ trials for each $x_o$.
\item For data from populations of size $N=100,000$, the procedure varied slightly between different parameter choices.  In Fig(2, main text) (left) and Fig(5, main text) we used $20,000$ trials each from $x_o \in \{10^{-4},5 \times 10^{-4}\}$.  In Fig(6, main text), we used $20,000$ trials from $x_o=2\times 10^{-4}$.  In Fig(2, main text)(right) we used $10,000$ trials from $x_o \in \{10^{-4} ,10^{-3} \}$\\
\end{itemize}

\section{Elaboration on A2$^*$}
\begin{figure}
\includegraphics[width=6in]{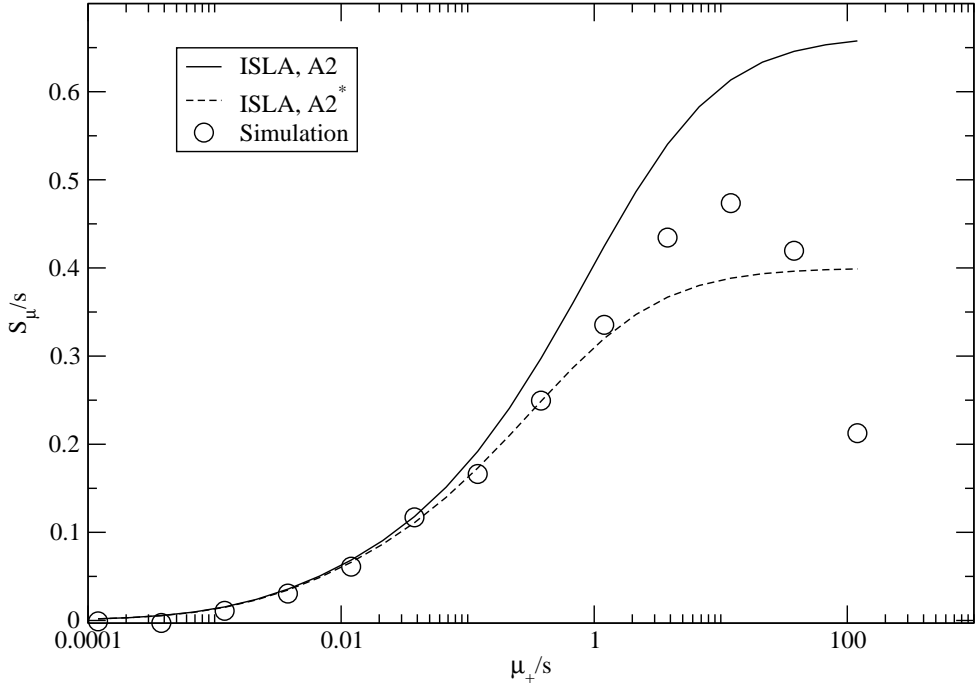}
\caption{The effect of using A$2^*$ instead of A2.  When $\mu_+/s \lesssim 1$, ISLA overestimates the results of simulations when it uses A2.  The opposite effect is observed if we instead make the assumption A$2^*$, which immediately kills the fraction (1-s) of advantageous mutants that are eventually lost to random drift.  This suggests that the error accumulated for $\mu_+/s \lesssim 1$ is due to the approximate manner in which ISLA treats these advantageous mutants.  Parameters are $N=5000$, $\mu_-=0$, $\alpha = .4$, $s=1/120$, $\delta = 0$.} 
\label{a2star_fig}
\end{figure}
As mentioned in the main text, A2 is somewhat awkward.  An alternative, which we call A2$^*$, it immediately kill advantageous mutations which are destined to eventually succumb to drift.  This approximation merely modifies a coefficient in Eq.\ref{back_eq}.  The effect is simply the transposition $\frac{\alpha_e}{1-\alpha_e} \rightarrow \alpha_e$.  In fact, we occasionally made this substitution in the text, when we anticipated that $\alpha_e \ll 1$.  Typical behavior of A2 relative to A2$^*$ is illustrated in Fig.\ref{a2star_fig}.  Even though A2$^*$ yields results that are arguably more accurate than those of A2, we preferred A2 in the main text because it nicely serves as an upper bound on mutator success. 

\section{Fixation and Loss Times when $\mu_-=0$ and $\mu_- > 0$}
\begin{figure}
\includegraphics[width=6in]{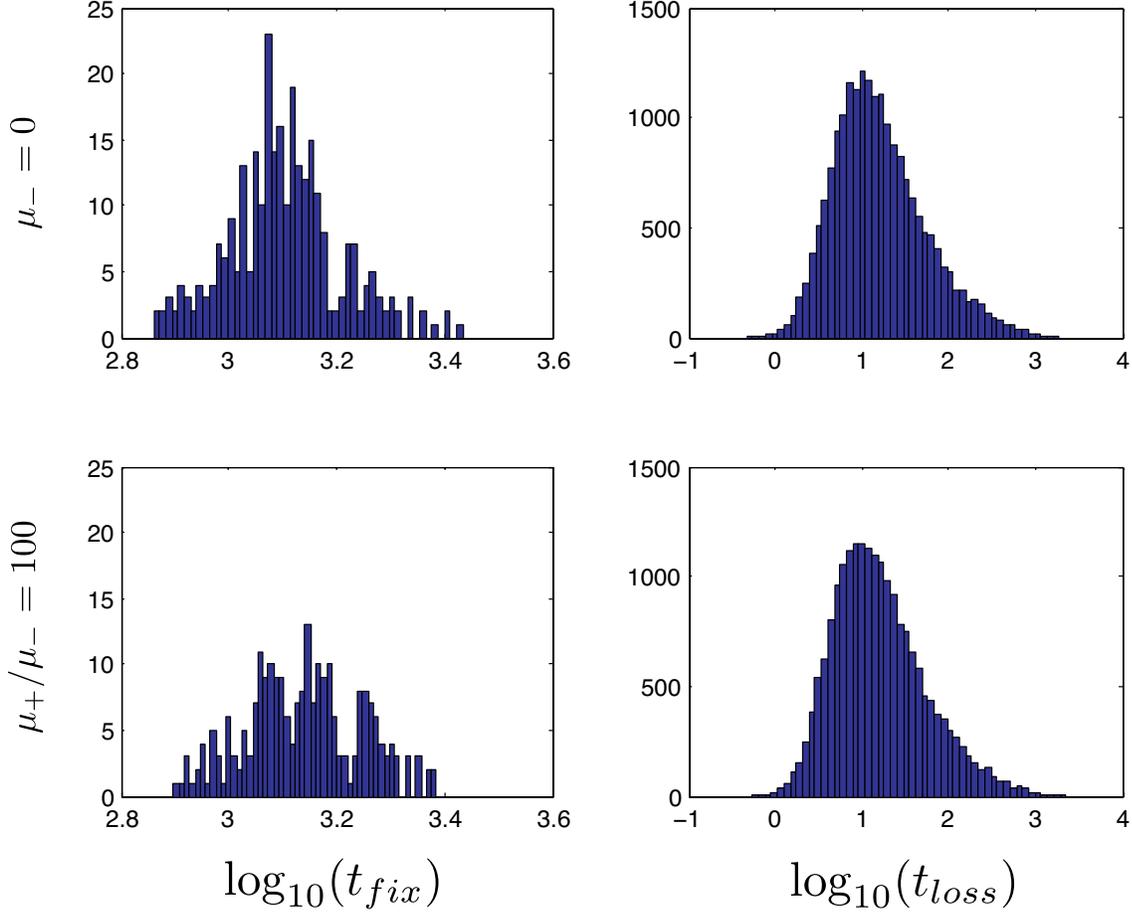}
\caption{The distributions of fixation and loss times for cases where $P_{fix} \approx 1 \%$.  The left (right) column shows the distribution of fixation (loss) times.  The top row corresponds to $\mu_- = 0$ and the bottom row to $\mu_+/\mu_-=100$.  Notice the logarithmic scale and the extremely long tails on the $t_{loss}$ distributions.  The two $t_{loss}$ distributions have the same mean $\bar{t}_{loss} \approx 40$ generations, which is of the same order as $\bar{t}_{drift} = \frac{\ln(Ns)}{s}\approx 92$ generations.  The $t_{fix}$ distributions have means $\bar{t}_{fix} \approx 1300$ generations ($\mu_- = 0$) and $\bar{t}_{fix}\approx 1400$ generations ($\mu_+/\mu_-=100)$.  Since $t_{sweep} \sim \frac{ln(Ns)}{s} \approx 800$ generations are required for an advantageous mutant to sweep the population, we see that $500-600$ generations passed before a beneficial mutant destined for fixation was generated.  Thus,
when mutator fixation occurs, such beneficial mutations are typically generated early compared to $\bar{t}_{mut} \equiv (\alpha s \mu_+ Nx_o)^{-1}= 3\times 10^4$ but late compared to $\bar{t}_{drift}$. $S_\mu$ is determined mostly by the probability that the mutator survives the long drift period and this is barely affected by wild-type beneficial mutant fixation events. Parameters are $N=10^5,s=1/120,\alpha=.4,x_o=10^{-4},\delta=0, \mu_+=10^{-3}$.  Note that the initial \textit{overall} mutation rate in the wild-type population is $100 \times$ that in the mutator subpopulation. \label{t_hists}}
\end{figure}

As mentioned in the main text, we do not fully understand why ISLA often fails in the weak-effect mutator regime.  To further explore this issue, in Fig.\ref{t_hists} we compared the distributions of fixation and loss times for $\mu_- = 0$ and $\mu_- > 0$.  We found very little difference in these distributions, suggesting that mutations in the wild-type subpopulation have only minor effects on the fixation process and apparently can be neglected.  The mechanism by which mutators succeed despite beneficial mutations in wild-type backgrounds is poorly understood and clearly deserves further attention in future work.  

\section{Simulations with Very Large s}
\begin{figure}
\includegraphics[width=6in]{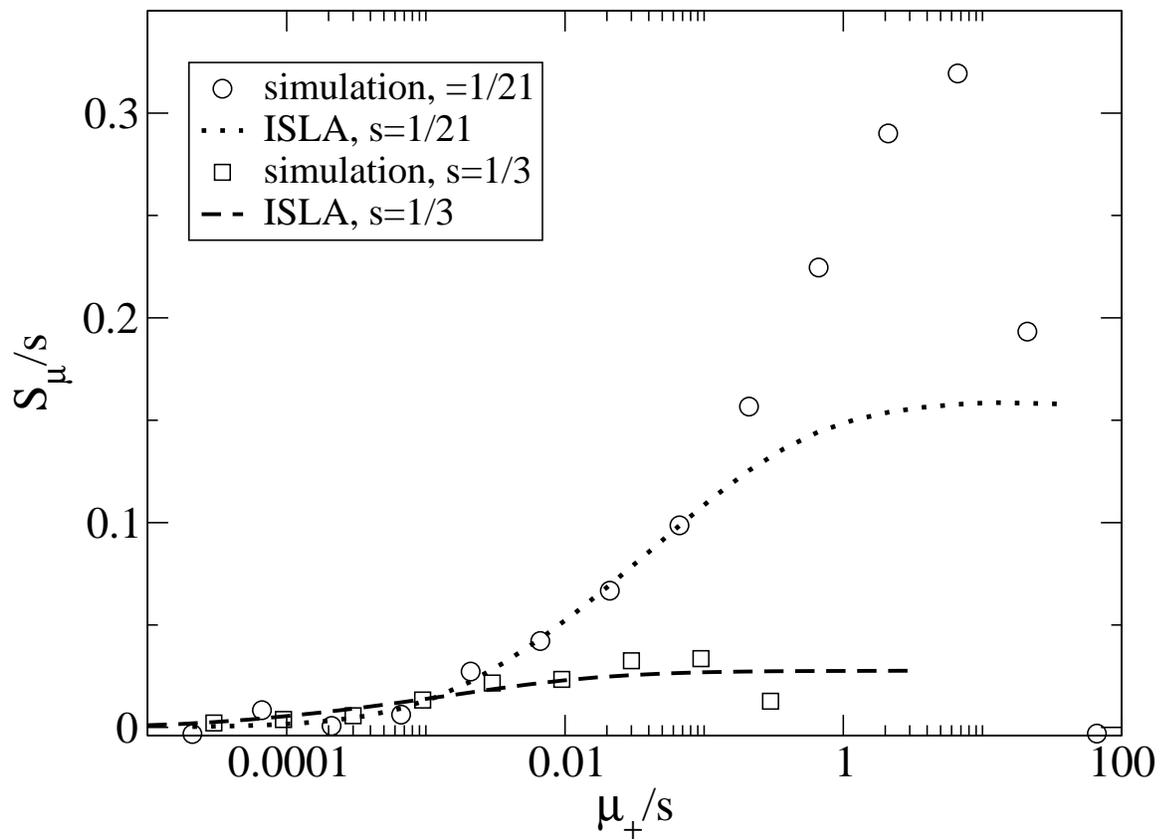}
\caption{Simulation data for very large $s$.  When $s=1/21$, ISLA greatly overestimates the the effect of mutations in wild-type backgrounds, whereas the agreement is much better when $s=1/3$.  Parameters are $N=1000$, $\mu_+/\mu_-=10, \alpha_e=.4, \delta=0$}  \label{v_large_s_fig}
\end{figure}

Fig.\ref{v_large_s_fig} shows that ISLA captures the effect of beneficial mutations in wild-type backgrounds only when $s$ is sufficiently large.  When $s=1/21$, ISLA greatly overestimates the the effect of mutations in wild-type backgrounds, whereas the agreement is much better when $s=1/3$.  We do not have a quantitative understanding of how large $s$ must be in order to achieve agreement.

\bibliographystyle{genetics}

\bibliography{sup_bib}

\begin{thebibliography}{16}
\expandafter\ifx\csname natexlab\endcsname\relax\def\natexlab#1{#1}\fi

\bibitem[{{\sc Boe} {\em et~al.\/}(2000){\sc Boe}, {\sc Danielsen}, {\sc
  Knudsen}, {\sc Petersen}, {\sc Maymann} {\em et~al.\/}}]{bo_mutator}
{\sc Boe, L.}, {\sc M.~Danielsen}, {\sc S.~Knudsen}, {\sc J.~Petersen}, {\sc
  J.~Maymann}, {\em et~al.\/}, 2000 The frequency of mutators in populations of
  escherichia coli.
\newblock Mutation Research {\bf 448}: 47--55.

\bibitem[{{\sc Desai} {\em et~al.\/}(2007){\sc Desai}, {\sc Fisher} and {\sc
  Murray}}]{d_fisher_short}
{\sc Desai, M.}, {\sc D.~Fisher}, and {\sc A.~Murray}, 2007 The speed of
  evolution and maintenance of variation in asexual populations.
\newblock Current Biology : 385--394.

\bibitem[{{\sc Eyre-Walker} and {\sc Keightley}(2007)}]{eyrewalker2007dfe}
{\sc Eyre-Walker, A.}, and {\sc P.~Keightley}, 2007 The distribution of fitness
  effects of new mutations.
\newblock Nature Reviews: Genetics {\bf 8}: 610.

\bibitem[{{\sc Gerrish} and {\sc Lenski}(1998)}]{gerrish1998fcb}
{\sc Gerrish, P.}, and {\sc R.~Lenski}, 1998 {The fate of competing beneficial
  mutations in an asexual population}.
\newblock Genetica {\bf 102}: 127--144.

\bibitem[{{\sc Hegreness} {\em et~al.\/}(2006){\sc Hegreness}, {\sc Shoresh},
  {\sc Hartl} and {\sc Kishony}}]{hegreness2006epi}
{\sc Hegreness, M.}, {\sc N.~Shoresh}, {\sc D.~Hartl}, and {\sc R.~Kishony},
  2006 {An equivalence principle for the incorporation of favorable mutations
  in asexual populations}.
\newblock Science {\bf 311}: 1615--1617.

\bibitem[{{\sc Imhof} and {\sc Schlotterer}(2001)}]{imhof2001fea}
{\sc Imhof, M.}, and {\sc C.~Schlotterer}, 2001 Fitness effects of advantageous
  mutations in evolving escherichia coli populations.
\newblock Proceedings of the National Academy of Sciences {\bf 98}: 1113.

\bibitem[{{\sc Keightley} and {\sc Eyre-Walker}(1999)}]{keightley1999tma}
{\sc Keightley, P.}, and {\sc A.~Eyre-Walker}, 1999 {Terumi Mukai and the
  riddle of deleterious mutation rates}.
\newblock Genetics {\bf 153}: 515--523.

\bibitem[{{\sc Kibota} and {\sc Lynch}(1996)}]{kibota1996egm}
{\sc Kibota, T.}, and {\sc M.~Lynch}, 1996 {Estimate of the genomic mutation
  rate deleterious to overall fitness in E. coli}.
\newblock Nature {\bf 381}: 694--696.

\bibitem[{{\sc Lenski} {\em et~al.\/}(1991){\sc Lenski}, {\sc Rose}, {\sc
  Simpson} and {\sc Tadler}}]{lenski1991lte}
{\sc Lenski, R.}, {\sc M.~Rose}, {\sc S.~Simpson}, and {\sc S.~Tadler}, 1991
  {Long-term experimental evolution in Escherichia coli. I. Adaptation and
  divergence during 2,000 generations}.
\newblock The American Naturalist {\bf 138}: 1315--1341.

\bibitem[{{\sc Orr}(2003)}]{orr2003dfe}
{\sc Orr, H.}, 2003 {The distribution of fitness effects among beneficial
  mutations}.
\newblock Genetics {\bf 163}: 1519--1526.

\bibitem[{{\sc Perfeito} {\em et~al.\/}(2007){\sc Perfeito}, {\sc Fernandes},
  {\sc Mota} and {\sc Gordo}}]{perfeito2007amb}
{\sc Perfeito, L.}, {\sc L.~Fernandes}, {\sc C.~Mota}, and {\sc I.~Gordo}, 2007
  Adaptive mutations in bacteria: High rate and small effects.
\newblock Science {\bf 317}: 813.

\bibitem[{{\sc Rozen} {\em et~al.\/}(2002){\sc Rozen}, {\sc de~Visser} and {\sc
  Gerrish}}]{rozen2002fef}
{\sc Rozen, D.}, {\sc J.~de~Visser}, and {\sc P.~Gerrish}, 2002 Fitness effects
  of fixed beneficial mutations in microbial populations.
\newblock Current Biology {\bf 12}: 1040--1045.

\bibitem[{{\sc Sniegowski} {\em et~al.\/}(1997){\sc Sniegowski}, {\sc Gerrish}
  and {\sc Lenski}}]{sniegowski1997ehm}
{\sc Sniegowski, P.}, {\sc P.~Gerrish}, and {\sc R.~Lenski}, 1997 {Evolution of
  high mutation rates in experimental populations of E. coli.}
\newblock Nature {\bf 387}: 659--661.

\bibitem[{{\sc Taddei} {\em et~al.\/}(1997){\sc Taddei}, {\sc Matic}, {\sc
  Godelle} and {\sc Radman}}]{taddei1997mop}
{\sc Taddei, F.}, {\sc I.~Matic}, {\sc B.~Godelle}, and {\sc M.~Radman}, 1997
  {To be a mutator, or how pathogenic and commensal bacteria can evolve
  rapidly}.
\newblock Trends in Microbiology {\bf 5}: 427--428.

\bibitem[{{\sc Wahl} and {\sc Gerrish}(2001)}]{wahl55pbm}
{\sc Wahl, L.}, and {\sc P.~Gerrish}, 2001 {The probability that beneficial
  mutations are lost in populations with periodic bottlenecks}.
\newblock Evolution {\bf 55}: 2606--2610.

\bibitem[{{\sc Wahl} {\em et~al.\/}(2002){\sc Wahl}, {\sc Gerrish} and {\sc
  Saika-Voivod}}]{wahl2002eip}
{\sc Wahl, L.}, {\sc P.~Gerrish}, and {\sc I.~Saika-Voivod}, 2002 {Evaluating
  the impact of population bottlenecks in experimental evolution}.
\newblock Genetics {\bf 162}: 961--971.

\end{thebibliography}

 \end{document}